\documentclass[aps,10pt,reprint]{revtex4-1}
\usepackage{mathrsfs}
\usepackage{graphicx}
\usepackage{epstopdf}
\usepackage{rotating}
\usepackage{courier}
\DeclareGraphicsExtensions{.ps}
\usepackage{amsmath,amssymb}
\usepackage[mathcal]{euscript}
\usepackage{color}
\usepackage{siunitx}
\newcounter{eqn}

\AppendGraphicsExtensions{.tif}

\makeatletter
\newcommand{\putindeepbox}[2][0.7\baselineskip]{{%
    \setbox0=\hbox{#2}%
    \setbox0=\vbox{\noindent\hsize=\wd0\unhbox0}
    \@tempdima=\dp0
    \advance\@tempdima by \ht0
    \advance\@tempdima by -#1\relax
    \dp0=\@tempdima
    \ht0=#1\relax
    \box0
}}
\makeatother

\begin{document}

\author{Jesse J. Lutz}
\email{jesse.lutz.ctr@afit.edu}
\thanks{This research was supported in part by an appointment to the Faculty Research Participation Program at
U.S. Air Force Institute of Technology (AFIT),
administered by the Oak Ridge Institute for Science and Education 
through an interagency agreement between the U.S. Department of Energy and AFIT.}
\email{jesse.lutz.ctr@afit.edu}
\affiliation{ORISE fellow residing at Department of Engineering Physics, Air Force Institute of Technology, Wright-Patterson Air Force Base, Ohio 45433 U.S.A.}
\author{Xiaofeng F. Duan}
\affiliation{Air Force Research Laboratory DoD Supercomputing Resource Center, Wright-Patterson Air Force Base, OH 45433 U.S.A.}
\author{Larry W. Burggraf}
\affiliation{Air Force Institute of Technology, Wright-Patterson Air Force Base, Ohio 45433 U.S.A.}

\title{A QM/MM equation-of-motion coupled-cluster approach for predicting
       semiconductor color-center structure and emission frequencies}

\begin{abstract}
Valence excitation spectra were computed for deep-center silicon-vacancy defects 
in 3C, 4H, and 6H silicon carbide (SiC) and comparisons were made with literature 
photoluminescence measurements. Optimizations of nuclear geometries surrounding 
the defect centers were performed within a Gaussian basis-set framework
using many-body perturbation theory or density functional theory (DFT) methods, 
with computational expenses minimized by a QM/MM technique called SIMOMM. 
Vertical excitation energies were subsequently obtained by applying excitation-energy, 
electron-attached, and ionized equation-of-motion coupled-cluster (EOMCC) methods, 
where appropriate, as well as time-dependent (TD) DFT, to small models including 
only a few atoms adjacent to the defect center. We consider the relative quality 
of various EOMCC and TD-DFT methods for 
(i) energy-ordering potential ground states differing incrementally in charge and multiplicity, 
(ii) accurately reproducing experimentally measured photoluminescence peaks, and 
(iii) energy-ordering defects of different types occurring within a given polytype.
The extensibility of this approach to transition-metal defects is also tested
by applying it to silicon-substituted chromium defects in SiC and comparing with measurements. 
It is demonstrated that, when used in conjunction with SIMOMM-optimized geometries, 
EOMCC-based methods can provide a reliable prediction of the ground-state charge and multiplicity, 
while also giving a quantitative description of the photoluminescence spectra, 
accurate to within 0.1 eV of measurement for all cases considered.

\end{abstract}

\keywords{Coupled-cluster theory, Equation-of-motion coupled-cluster methods, 
nitrogen vacancy defect, Excited electronic states, 
Excitation spectra, Silicon carbide defects} 

\maketitle

\section{Introduction}
\label{sec_intro}

Certain point defects in wide-band-gap semiconductors have been identified
as promising candidates for use as qubits in quantum computing, communication, and sensing applications \cite{NMRs2008}.
A well-known example is the nitrogen-vacancy (NV) color center in diamond, 
which harbors an anionic electronic structure [(NV)$^-$] with well-defined $S=1$ spin states 
that have been initialized and coherently manipulated using optical or microwave radiation \cite{DMDpr2013}.
The resulting stimulated emission, occurring between the $^3$A$_2$ ground state and the $^3$E excited state,
produces a tunable photoluminescence, polarized according to an applied external magnetic field.
The demonstration of long spin-coherence times at room temperature established (NV)$^-$ centers
as one of the most stable, efficient, high-quality single-photon sources known \cite{CJInm2014,CFAnm2015,WLRnm2015}.
However, diamond has inherent engineering limitations and, as a result, defects with similar properties 
are being eagerly sought out, both in other solid-state materials \cite{AETnp2016} and in nano-materials \cite{HHMnp2017}.

The most closely related material to diamond in terms of $sp^3$ bonding is silicon carbide (SiC),
and its anionic silicon-vacancy (V$_{\mathrm{Si}}^-$) defects are arguably better qubit candidates
than the diamond (NV)$^-$ defect. The SiC V$_{\mathrm{Si}}^-$ defects, 
characterized by a $^4$A$_2 \rightarrow {}^4$E transition \cite{SDEprb2016},
have several superior properties: they exhibit no luminescence intermittency or ‘blinking’ \cite{BGNnn2010,WLRnm2015},
they are a half-integer $S=\frac{3}{2}$ spin (thus Kramers theorem holds) \cite{DDFprb2012,LVZapl2015},
and they are intrinsic defects, which do not require doping 
and can therefore be more easily created
(e.g., using a transmission electron microscope \cite{SEDdrm2002},
a focused ion beam \cite{WZZacsp2017}, ion implantation \cite{WZZpra2017}, etc.),
as reported in Ref.\ \cite{RWNnl2017}, 
where a scalable array of single silicon vacancy centers was realized \cite{RWNnl2017}.
Furthermore, the bulk material properties of SiC make it more amenable than diamond 
to high-voltage, high-power, and high-temperature applications
and it is also more promising as a long-term candidate material due to its physical durability \cite{KCbook2014},
engineering flexibility \cite{FBCnc2013}, and increasingly inexpensive manufacturing cost \cite{HRRnrel2017,WCQdt2017}.
One disadvantage is that the optically detected magnetic resonance of V$_{\mathrm{Si}}^-$ SiC 
has a lower visibility compared to that of the (NV)$^-$ center in diamond, but this too
is being overcome \cite{NWNa2017}.

Diamond (NV)$^-$ and SiC V$_{\mathrm{Si}}^-$ defects emit in a region of the infrared
which is non-ideal for utilizing existing singlemode fiber-optic infrastructure.
Recent telecommunications systems use wavelength-division multiplexing, 
which can use the full range of wavelengths between 1260 and 1670 nm (or 0.74 and 0.98 eV), 
and other popular multimode and singlemode fiber implementations operate 
using wavelengths of 850, 1300, and 1550 nm (or 0.80, 0.95, and 1.46 eV),
with the latter being associated with the transmission-optimal so-called C-band.\cite{OHMapl2017}
Consequently, several other common SiC defects are also under consideration, 
including the neutral divacancy \cite{CJInm2014,SFKnc2016}, nitrogen vacancies \cite{CvBCa2017,vBCmrsc2017},
and anti-site vacancies \cite{TNIprl2006,SIAprb2015}. More promising still 
is the prospect of doping with transition- or heavy-metal elements \cite{KDWprb2017},
although it is unclear at the outset which implants will emit the desired wavelengths.

Difficulties are often encountered when using spectroscopic techniques to distinguish 
between different types of vacancies or screen for specific properties 
across a series of substitutional defects, presenting a great opportunity
for computational modeling. Modeling photoluminescence spectra requires
determination of the nuclear geometry of the solid-state defect, followed by
generation of accurate wave-functions for both the ground and excited states.
Oftentimes acceptable geometries can be obtained for weakly correlated materials
using density functional theory (DFT) methods in a plane-wave basis, and its
computational scaling, usually $\mathscr{N}^3$--$\mathscr{N}^4$ with the system size $\mathscr{N}$,
offers a realtively inexpensive framework. Unfortunately, problems arise when extending DFT
to excited states through the time-dependent (TD) DFT formalism
(see, e.g., Refs.\ \cite{MGarpc2004,DHcr2005,LJijqc2013,AJcsr2013,MarXiv} for reviews).
Alternatively, studies conducted in a plane-wave basis may apply the many-body $GW$ approximation \cite{AGrpp1998},
sometimes even in conjunction with DFT-optimized structures \cite{BRprb2011},
in order to gain access to band structure or excited states.
The $GW$ approximation provides much more accurate results, but it also has well-known fundamental limitations:
to name a few, it suffers from self-consistency errors and the route toward an exact theory is unclear.
 
Among the most accurate general-purpose {\it ab initio} methods available 
are those based on the single-reference coupled-cluster (SRCC) theory for ground states
and its equation-of-motion (EOM) CC extension to excited states.
These methods are size-consistent and systematically improvable,
but, despite recent progress toward reducing the expense
of plane-wave basis SRCC/EOMCC implementations \cite{LGjcp2016,HTGjcp2017},
their computational scaling remains intractable for solids.
An acceptable alternative for geometry optimizations is provided by
the related second-order many-body perturbation theory [MBPT(2)],
which has a noniterative $\mathscr{N}^5$ scaling, but for an accurate
treatment of excitation energies EOMCC-based methods are needed.
The most basic EOMCC methods, including only single and double excitations,
require steep iterative $\mathscr{N}^5$--$\mathscr{N}^6$ scalings,
and this makes treatment of even a single unit cell very taxing 
in terms of the required CPU cycles. Meanwhile explicit treatment 
of a supercell model with Gaussian-based {\it ab initio} methods 
will be impossible for many years to come, even with modernized codes.

Here a two-fold strategy is used to minimize the computational expense 
associated with the aforementioned accurate computational methodologies. 
The first step is to partition the geometry optimization into a small 
group of atoms significantly perturbed by introduction of the defect site, 
and a comparatively very large group of atoms whose environment is 
unchanged by introduction of the defect. The former will be treated
using high-level quantum-based methods, while the latter will be treated
using low-level classical-based methods.
The second step is to exploit the highly-localized nature of the associated
defect photoluminescence by applying accurate excited-state many-body methods 
to small model systems, e.g., only those few atoms directly adjacent to the defect.
For over 50 years the local nature of excitations in defect solids 
has been used to develop approximate methods in which the total system is 
is subdivided into a defect subspace and a complementary crystalline region,
so this is not a novel proposition.\cite{Kbook1968}

The first step is realized by utilizing the surface integrated molecular-orbital 
molecular-mechanics (SIMOMM) method of Shoemaker et al. \cite{Shoemaker}, which falls
into the general class of quantum-mechanics/molecular-mechanics (QM/MM) hybrid methods.
The SIMOMM framework imposes a less rigid treatment of the capping atoms 
than its predecessor, the IMOMM model of Maseras and Morokuma \cite{Morokuma},
and this reduces artificial strain imposed on the QM structure.
SIMOMM was originally developed for the study of surface chemical systems,
and by now its utility has been demonstrated repeatedly for describing chemistry
on Si and SiC surfaces \cite{CGjacs1999,LBSapl2000,JCGjpcb2001,
CGjacs2002,CLEjacs2002,TGjcp2003,RGjpcb2004,JGjacs2005,LCGjpcb2005,LCGjacs2005}.
SIMOMM geometry optimizations are perfomed under the Born-Oppenheimer approximation,
which in this case is based on extremely robust assumptions \cite{LHjms2016}.
The current study is the first attempt at applying SIMOMM to describe
nuclear geometries of deep-center defects in semiconductors. Note
that for metallic defects other, more appropriate, QM/MM methods exist 
(see, e.g., Refs. \cite{CLEprb2005} and \cite{LLCmsmse2007}).

The second step of the abovementioned procedure is also very challenging
due to the nature of the electronic excitations of interest. 
Excitation energies of open-shell systems with high spacial symmetry 
are notoriously difficult to describe, and, as a result, we employ 
the electron-attached (EA) and ionized (IP) EOMCC methods \cite{NBjcp1995a,NBjcp1995b}.
These methods have been shown to be particularly accurate for describing
both ground and excited states of odd-electron, open-shell molecules.
In these schemes the ($N\pm 1$)-electron systems of interest 
are formed through application of an electron-attaching or ionizing operator 
to the correlated ground-state reference of a related $N$-electron system,
obtained using the SRCC approach. Alternatively, if even-electron, open-shell
states are desired, they can be described by the excitation-energy (EE) EOMCC method,
where the usual particle-conserving operator is applied to the same correlated
$N$-electron reference.
This framework allows for orthogonally spin-adapted and systematically improvable
calculations of the ground and excited states of $N$- and ($N\pm 1$)-electron
systems mutually related by an $N$-electron correlated reference function.

Electronically excited states dominated by one-electron transitions,
particularly those that correspond to one-electron transitions from 
non-degenerate doubly-occupied molecular orbitals (MOs) to a singly-occupied 
molecular orbital (SOMO), can be accurately described
by the basic EE-, EA-, or IP-EOMCCSD approaches. Meanwhile, electronic
transitions characterized by two- or other, more complicated, many-electron processes, 
require higher-than-double excitations in order to obtain reliable results. 
The expense of such calculations usually limits their applicability to the smallest systems,
but larger systems can be efficiently treated 
using the active-space EE-, EA-, and IP-EOMCC 
variants \cite{OAjcp1991,OAjcp1992,OAirpc1993,POAjcp1993},
such as those including active-space triples, i.e., 
EE-, EA-, and IP-EOMCCSDt \cite{GPWjcp2005,GPjcp2006}.
Here a strategically-chosen small subset of orbitals is considered 
that captures the largest contributions from triple excitations.

The performance of both the basic EE-, EA-, and IP-EOMCCSD methods
and the active-space EA- and IP-EOMCCSDt approaches
are tested here for their ability to describe solid-state SiC defect emission frequencies.
The active-space methods have already been applied to small open-shell 
molecules \cite{GPWjcp2005,GPjcp2006,HPLps2011} 
and ionic transition-metal complexes \cite{EPLcp2012}, where it was demonstrated
that they can provide an accurate treatment as compared to calculations
employing a full treatment of triple excitations. The current study 
is is the first application to deep-center defects, where the resulting
EOMCC excitation energies can be directly compared to photoluminescence measurements.
Some TD-DFT calculations are included for comparison, but benchmarking various
functionals is outside of the scope of this work so we use those deemed
optimal for Si$_n$C$_m$ ($n\leq m\leq 12$) molecules in Refs.\ \cite{BLJjcp2016} 
and \cite{LDBjcp2017} (see Sect.\ \ref{sec_comp} for further details). 

The goal of the present work is to develop an accurate procedure for 
describing emission frequencies, and we use as benchmarks the available
photoluminescence spectra for the V$_{\mathrm{Si}}^-$ defects in 4H- and 6H-SiC,\cite{WMCprb2000}
and also the chromium silicon-substitutional defect in SiC.\cite{KDWprb2017}
Two distinct Si-vacancy sites exist in 4H-SiC, the $k$-site and the $h$-site,
and each exhibit distinct signature photoluminescence in the infrared.
Meanwhile, three distinct sites exist in 6H-SiC, the $k1$- $k2$- and $h$-sites,
each of which also emit in the infrared. The V$_{\mathrm{Si}}^-$ defect in
3C-SiC has not been widely reported as it potentially undergoes low-temperature 
annealing,\cite{SKpb1993,IKOpss1997} but there is evidence it lies in
the same range as V$_{\mathrm{Si}}^-$ defects in the other polytypes (1.3--1.4 eV).\cite{OSMprb2003}
The measured emission frequencies of each of these six defects all fall within
0.1 eV of one another, but selective resonant optical excitation is still possible,
as the spectral linewidths can be as small as 2 $\mu$eV \cite{DFKprl2012}.
A useful theory will be able to predict each frequency to an accuracy within 0.1 eV,
while also giving the correct qualitative energy-ordering of closely-spaced emission frequencies.

The structure of this paper is as follows: in Sect.\ \ref{sec_theo}
the basic theory of the EOMCC methods is presented, while in Sect.\ \ref{sec_comp}
specific details are given about how the computations were performed.
Sect.\ \ref{sec_randd} details investigations of SIMOMM convergence,
qualitative and quantitative energy-ordering of states with incremental 
changes in charge and multiplicity, and comparison of computed excitation energies
with photoluminescence measurements for various defect sites.
Conclusions and directions for future research are discussed in Sect.\ \ref{sec_conc}.

\section{Theory}
\label{sec_theo}

\begin{figure}
\includegraphics[width=0.45\paperwidth]{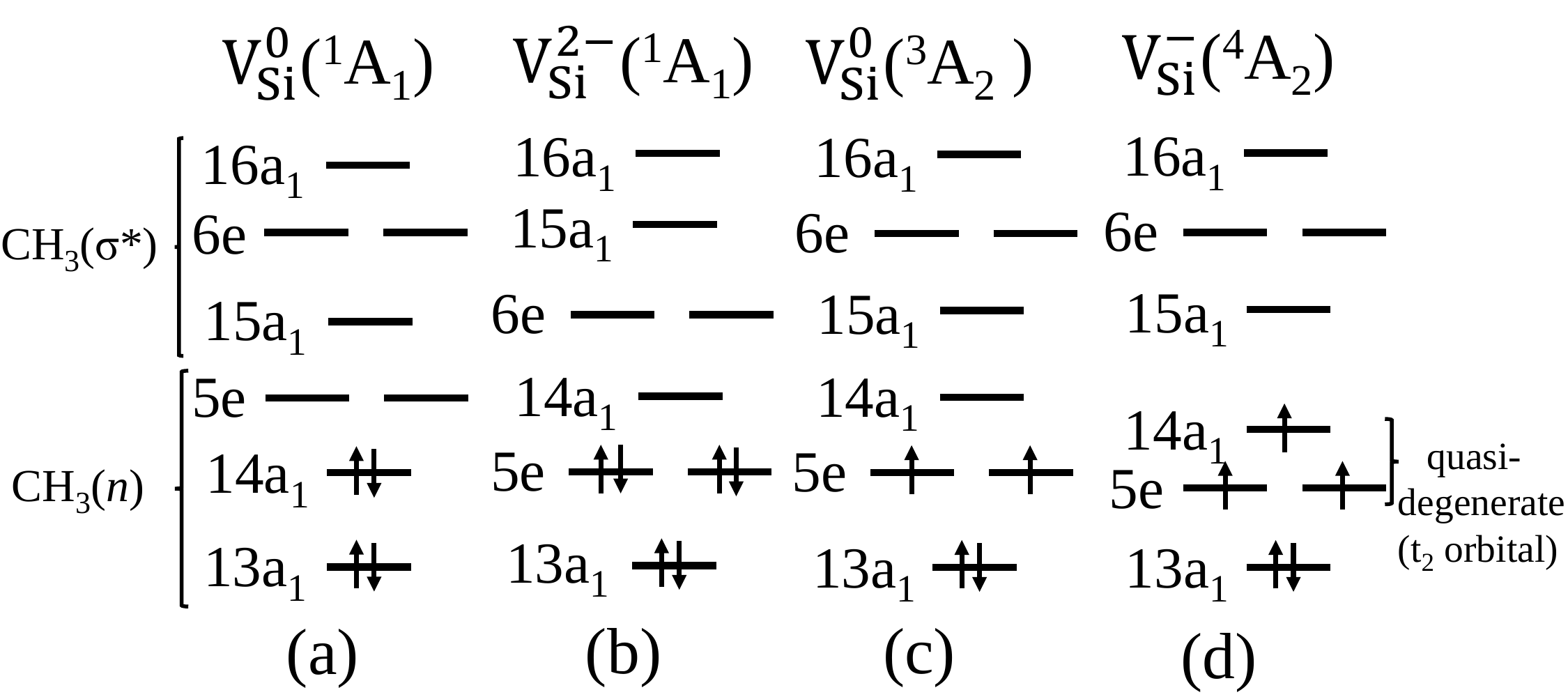}
\caption{\label{fig_ipm} Single-particle representations of several potential ground states
                         for V$_{\mathrm{Si}}$ and their qualitative valence-orbital energy-orderings
                         and occupations.}
\end{figure}

In this section we provide an overview of the EOM-CC methods 
used to describe the excitation energies of the SiC color centers.
The EE-, EA-, and IP-EOMCC theories will be employed to consider 
relative energies of ground and excited states with varying charge and multiplicity. 
These EOMCC-based methods have several advantages over standard DFT and TD-DFT:
they produce spin-adapted odd-electron states, they are systematically improvable,
and they can be used to generate $N$- and ($N\pm1$)-electron states 
of various multiplicities from a common correlated $N$-electron reference function. 

In general, a wave function $|\Psi_{\mu}\rangle$, corresponding to state $\mu$ of interest,
is expressed by applying a linear excitation operator $R_{\mu}$
to the ground-state SRCC wave function, 
\begin{equation}
|\Psi_{\mu}\rangle = R_{\mu} |\Psi^{(N)}_0\rangle,
\label{eq_wf}
\end{equation}
where $|\Psi_0\rangle=e^T|\Phi\rangle$ is the CC ground state wave function
formulated from the many-body cluster operator $T$ and $|\Phi\rangle$ 
is for the EOMCC methods in this work always given by 
the restricted Hartree-Fock (RHF) wave function, $|\Phi\rangle = |\Phi^{\mathrm{RHF}}\rangle$.
By choosing as a starting point an $N$-electron reference CC function, $|\Psi^{(N)}_0\rangle$, 
we are able to maintain commutation relations with the $S^2$ and $S_z$ operators throughout.
To access both $N$- and ($N\pm 1$)-electron states, the linear excitation operator $R_{\mu}$ 
must be either particle-conserving, $R_{\mu}=R_{\mu}^{(N)}$, or particle-nonconserving, $R_{\mu}=R_{\mu}^{(N\pm 1)}$, 
respectively, for the resulting state to be a spin-eigenfunction of the Hamiltonian.

In the particle-conserving EE-EOMCC theory, excited state energies and wave functions
are obtained for an $N$-electron system by applying in Eq.\ \ref{eq_wf} 
a linear excitation operator $R^{(N)}_{\mu}$ of the form
\begin{align}
R^{(N)}_{\mu} & = R_{\mu,0} + R_{\mu,1} + R_{\mu,2} + \ldots \\
              & = r_{\mu,0} + \sum_{\substack{ a \\ i}} r^i_a a^a a_i 
                            + \sum_{\substack{ ab \\ ij}} r^{ij}_{ab} a^a a^b a_j a_i + \ldots \nonumber
\label{eq_ree}
\end{align}
where $i$,$j$,$\ldots$($a$,$b$,$\ldots$) are the occupied (unoccupied) orbitals in $|\Psi^{(N)}_0\rangle$,
$a^p$($a_p$) are the creation (annihilation) operators associated with the spin-orbital basis set ${|p\rangle}$
used in the calculations and $r^i_a$, $r^{ij}_{ab}$,$\ldots$ are the excitation amplitudes
defining the many-body components of $R^{(N)}_{\mu}$, determined by diagonalizing the
similarity-transformed Hamiltonian $\bar{H} = e^{-T^{(N)}}He^{T^{(N)}}$ 
resulting from the ground-state $N$-electron CC calculations.

In the particle-nonconserving EA- and IP-EOMCC approaches, ground- and excited-state wave functions
are obtained corresponding to states with ($N-1$)- or ($N+1$)-electron open-shell systems, respectively.
The corresponding electron-attaching and ionizing operators, $R^{(N+1)}_{\mu}$ and $R^{(N-1)}_{\mu}$,
respectively, entering Eq.\ \ref{eq_wf} are defined as
\begin{align}
R^{(N+1)}_{\mu} & = R_{\mu,1p} + R_{\mu,2p-1h} + R_{\mu,3p-2h} + \ldots \\
                & = \sum_a r_a a^a + \sum_{\substack{a<b \\ j}} r^j_{ab} a^a a^b a_j 
                                   + \sum_{\substack{a<b<c \\ j>k}} r^{jk}_{abc} a^a a^b a^c a_k a_j + \ldots \nonumber
\label{eq_rea}
\end{align}
and 
\begin{align}
R^{(N-1)}_{\mu} & = R_{\mu,1h} + R_{\mu,2h-1p} + R_{\mu,3h-2p} + \ldots \\
                & = \sum_a r^i a_i + \sum_{\substack{i>j \\ b}} r_b^{ij} a^b a_j a_i  
                                   + \sum_{\substack{i>j>k \\ b<c}} r_{bc}^{ijk} a^b a^c a_k a_j a_i + \ldots \nonumber
\label{eq_rip}
\end{align}
where $r^i$, $r^{ij}_b$, $r^{ijk}_{bc}$, $\ldots$ and $r_a$, $r_{ab}^j$, $r_{abc}^{jk}$, $\ldots$ 
are the corresponding electron-attaching or ionizing amplitudes defining the relevant 
$1h$, $2h-1p$, $3h-2p$, $\ldots$ or $1p$, $2p-1h$, $3p-2h$, $\ldots$ components of 
$R^{(N+1)}_{\mu}$ and $R^{(N+1)}_{\mu}$, respectively, determined by diagonalizing the 
similarity tranformed Hamiltonian in the appropriate sector of the Fock space.

Active-space approaches represent a practical way to account for higher-than-doubly 
excited clusters in the CC and EOMCC equations \cite{OAjcp1991,OAjcp1992,OAirpc1993,POAjcp1993,Pmp2010}.
The idea is to sub-partition of the one-electron basis of occupied and unoccupied spin-orbitals
into (i) core or inactive occupied spin-orbitals, designated as {\bf i},{\bf j},$\ldots$, 
(ii) active occupied spin-orbitals, designated as {\bf I},{\bf J},$\ldots$, 
(iii) active unoccupied spin-orbitals, designated as {\bf A},{\bf B},$\ldots$,
and (iv) virtual or inactive unoccupied spin-orbitals, designated as {\bf a},{\bf b},$\ldots$.
After dividing the available orbitals into one of these four categories,
only active orbitals are used to define the active-space component of 
the EE, EA, or IP operators $R_{\mu}^{(N)}$, $R_{\mu}^{(N+1)}$, or $R_{\mu}^{(N-1)}$, respectively.
As an example, the active-space EA-EOMCCSDt\{$N_u$\} approach using
$N_u$ active unoccupied orbitals is obtained by replacing the $3p-2h$ component $R_{\mu,3p-2h}$
of the electron attaching operator $R_{\mu}^{(N+1)}$, Eq.\ \ref{eq_rea}, by
\begin{equation}
r_{\mu,3p-2h} = \sum_{\substack{j>k \\ {\bf A}<b<c}} r_{{\bf A}bc}^{jk} a^{\bf A} a^b a^c a_k a_j.
\label{eq_as}
\end{equation}
Assuming a small active space is chosen, there will be relatively few amplitudes $r_{{\bf A}bc}^{jk}$
defining $r_{\mu,3p-2h}$ in Eq.\ \ref{eq_as}, and they will not be much more expensive to compute
than the remining $1p$ and $2p-1h$ amplitudes $r_a$ and $r_{ab}^{ j}$ that enter the ($N+1$)-electron
wave functions of the active-space EA-EOMCCSDt\{$N_u$\} approach. 
The IP-EOMCCSDt\{$N_o$\} active-space method is formulated in an analogous way.

\begin{figure}
\includegraphics[width=0.4\paperwidth]{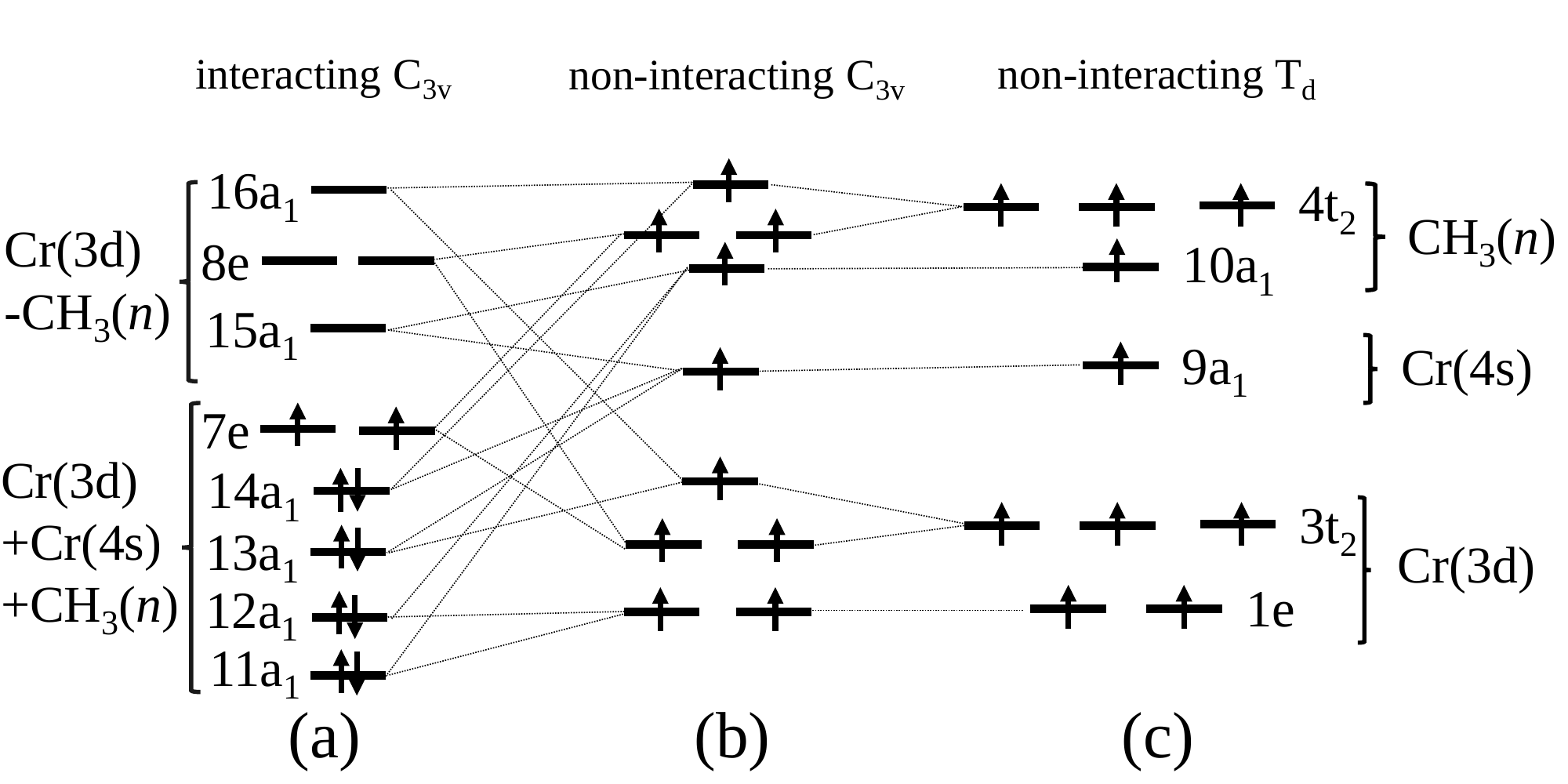}
\caption{\label{fig_tmmo} Schematic representation of the valence orbital energy levels
for Cr(CH$_3$)$_4$ assuming C$_{\mathrm{3v}}$ and T$_{\mathrm{d}}$ symmetries
with the former also considered with and without orbital mixing.}
\end{figure}

\section{Computational details}
\label{sec_comp}

\begin{figure*}
\includegraphics[width=0.25\paperwidth]{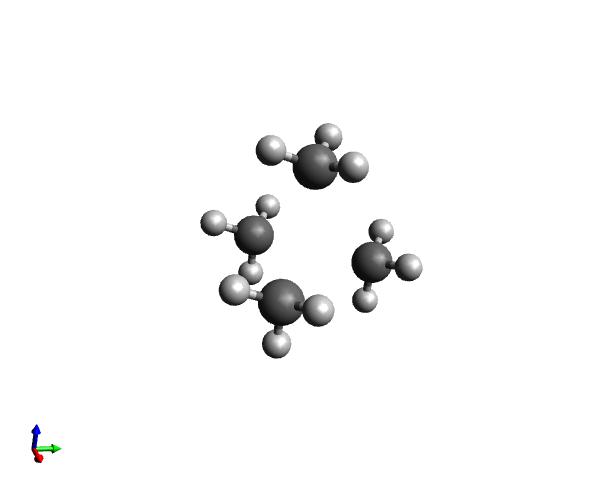}
\includegraphics[width=0.25\paperwidth]{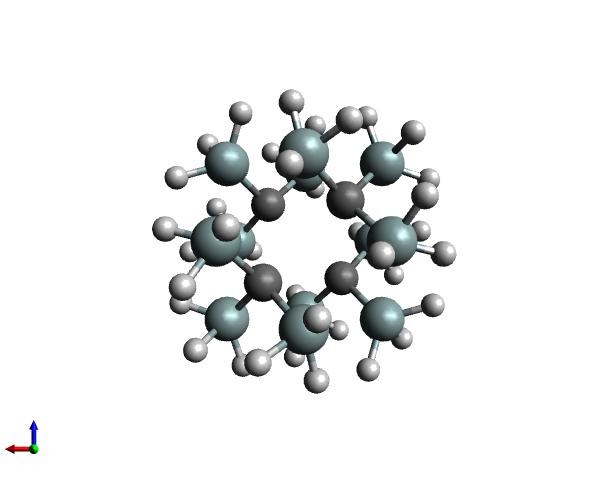}
\includegraphics[width=0.25\paperwidth]{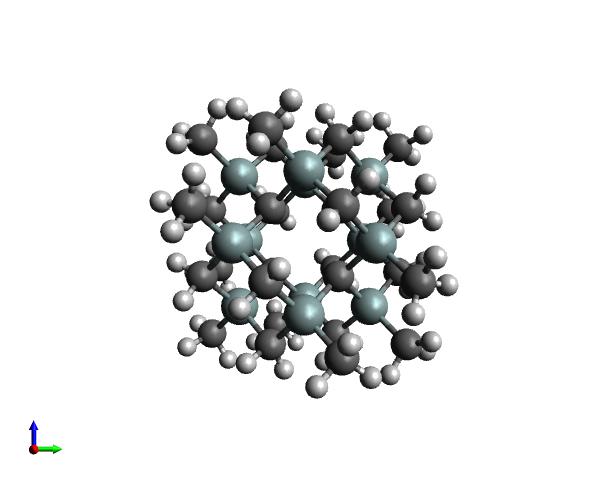} \\
(a) \hspace{2in} (b) \hspace{2in} (c)
\caption{\label{fig_mod} Atomic configurations of the QM models
corresponding to the C$_4$H$_{12}$ (a), C$_4$Si$_{12}$H$_{36}$ (b), and
C$_{40}$Si$_{12}$H$_{120}$ h-center V$_{\mathrm{Si}}^-$ defect in 4H-SiC.}
\end{figure*}

Here we outline necessary details for the calculations reported in Sect.\ \ref{sec_randd}.
Supercells with perfect lattice geometries were generated using the VESTA package \cite{vesta2011}.
Perimeter C atoms were capped with hydrogens, Si atoms were capped with CH$_3$ groups,
and defect sites were embedded into the cluster models using Avagadro and Avagadro 2 \cite{HCLjc2012}. 
The resulting geometries were converted to the proper format for the Tinker package \cite{tinker1987}
using OpenBabel \cite{BBJjc2011} and for the GAMESS package \cite{gamess1993,gamess2005}
using MacMolPlt \cite{BGjmgm1998}. 

Geometry optimizations are accellerated significanty by 
parallel implementations leveraging analytic gradients.
In a previous study we have investigated the relative performance of
a variety of DFT, MBPT(2), and high-level CC methods for
producing geometries of several Si$_n$C$_m$ ($n\leq m\leq 12$) molecules \cite{BLJjcp2016},
which are expected to exhibit similar many-body physics to defects in solid-state SiC.
The previous study found that MBPT(2) and DFT with the M11 functional
were good alternatives, which could closely reproduce the geometries
predicted by high-level SRCC methods. 
Since it is known in advance that the V$_{\mathrm{Si}}^{-}$ defects
have ${}^4$A$_1$ ground state, unrestricted (U) self-consistent field
variants were employed where appropriate, i.e., UMPBT(2) and UM11.
Unlike the open-shell coupled-cluster codes, both the UMBPT(2) and UM11 
methods have parallel analytic gradients implemented in GAMESS \cite{AGjpca2004},
and consequently they are used here for the QM portion of optimizations.
Restricted open-shell HF (ROHF) references were also tested,
but convergence problems were encountered.

The T$_{\mathrm{d}}$ (C$_{\mathrm{6v}}$) symmetry of bulk 3C- (4H- and 6H-) SiC are lowered 
to C$_{\mathrm{3v}}$ in the presence of vacancy defects such as V$_{\mathrm{Si}}^{-}$.
Unfortunately, the SIMOMM approach does not currently utilize the spacial symmetry of Abelian groups,
as is otherwise fully implemented in GAMESS. The SIMOMM-optimized structures reported here 
retained an approximate C$_{\mathrm{3v}}$ symmetry, but to facilitate comparisons with
other results found in the literature we had to trace back C$_{\mathrm{3v}}$ orbital labelings.
Due to the lowered symmetry, an otherwise degenerate E excited state in C$_{\mathrm{3v}}$ symmetry
had slightly different energies; in such cases we report the average of the two energy levels, 
which usually differed by a small amount (in many cases 1--5 meV). Consequently, 
all reported energy values are rounded to the nearest 0.01 eV, except when
more decimals are needed for qualitative discussions.

Excited-state calculations were performed using the EOMCC and TD-DFT approaches,
The B3LYP functional was chosen since it performed particularly well 
for Si$_n$C$_m$ ($n\leq m\leq 12$) clusters in Refs.\ \cite{LDBjcp2017} and \cite{Boyd}.
For TD-DFT calculations of open-shell species, only unrestricted (U) Kohn-Sham (KS) determinants were employed, 
as it has been advocated by Pople, Gill, and Handy that ROHF KS determinants 
should be avoided whenever possible \cite{PGHijqc1995}.
These methods were applied to SIMOMM-optimized geometries including only 
the four carbon atoms immediately surrouding the defect with three 
capping hydrogen atoms each. 
Ground states are labeled with an X, while roman numerals label
excited states of each symmetry, starting with 1. 
All computed excitation energies reported
here are vertical, which is expected to be a good approximation 
for solid-state photoluminescence phenomena.

For EA-EOMCCSDt calculations on the V$_{\mathrm{Si}}^{-}$ defect, 
a neutral CCSD reference [Fig.\ \ref{fig_ipm}(a)] was used with active-space orbitals 
chosen as 5e and 15a$_1$ 
in order to construct the corresponding quartet state [Fig.\ \ref{fig_ipm}(d)].
For the IP-EOMCCSDt calculations a doubly-anionic reference
[Fig.\ \ref{fig_ipm}(b)] was used with active-space orbitals 
chosen as 4e, 5e, 12a$_1$, and 13a$_1$ 
in order to construct the corresponding quartet state [Fig.\ \ref{fig_ipm}(d)].
Tests including more active-space orbitals did not have 
a significant effect on the excitation energies of interest.
The EE-EOMCCSDt method is not currently available in GAMESS, so all EE-EOMCC calculations
include singles and doubles only.
For EE-EOM-CCSD calculations a neutral CCSD reference [Fig.\ \ref{fig_ipm}(a)]
was used in order to construct the corresponding triplet state [Fig.\ \ref{fig_ipm}(c)].

\begin{table*}
\caption{\label{tab_simomm} Average distances ($\bar{R}$) between 
the h-center 4H-SiC V$_{\mathrm{Si}}^{-}$ defect and the four surrounding atoms, 
with convergence of the \% difference observed for various aspects of the SIMOMM model.}
\begin{tabular}{lllcccccc}
\hline
\hline
\multicolumn{3}{c}{SIMOMM model specifications}  &&           && \multicolumn{3}{c}{$\Delta\bar{R}$ (\% difference)\footnotemark[1]} \\
\cline{1-3}\cline{6-9}
method     & supercell  & QM model               && $\bar{R}$(\AA)&&  QM model  &  MM model  & basis set  \\
\hline
UMBPT(2)/STO-3G&  None      & C$_4$H$_{12}$          &&  1.814    &&            &            &            \\
UMBPT(2)/STO-3G&  None      & C$_{40}$Si$_{12}$H$_{120}$&& 2.321  &&    5.9     &            &            \\
UMBPT(2)/STO-3G&  4x4x1     & C$_4$H$_{12}$          &&  1.967    &&            &    8.1     &            \\
UMBPT(2)/STO-3G&  4x4x1     & C$_4$Si$_{12}$H$_{36}$ &&  2.038    &&    3.5     &    7.1     &            \\
UMBPT(2)/STO-3G&  4x4x1     & C$_{40}$Si$_{12}$H$_{120}$&& 2.076  &&    1.9     &   11.1     &            \\
UMBPT(2)/STO-3G&  8x8x2     & C$_4$H$_{12}$          &&  1.936    &&            &    1.6     &            \\
UMBPT(2)/STO-3G&  8x8x2     & C$_4$Si$_{12}$H$_{36}$ &&  2.005    &&    3.5     &    1.6     &            \\
UMBPT(2)/STO-3G&  8x8x2     & C$_{40}$Si$_{12}$H$_{120}$&& 2.040  &&    1.7     &    1.8     &            \\
\hline
UMBPT(2)/CCD   &  None      & C$_4$H$_{12}$          &&  1.856    &&            &            &    2.3     \\
UMBPT(2)/CCD   &  None      & C$_4$Si$_{12}$H$_{36}$ &&  3.895    &&   46.5     &            &   56.1     \\
UMBPT(2)/CCD   &  4x4x1     & C$_4$H$_{12}$          &&  2.013    &&            &    8.1     &    2.3     \\
UMBPT(2)/CCD   &  4x4x1     & C$_4$Si$_{12}$H$_{36}$ &&  2.338    &&   14.9     &   50.0     &   13.7     \\
UMBPT(2)/CCD   &  8x8x2     & C$_4$H$_{12}$          &&  1.964    &&            &    2.5     &    1.4     \\
UMBPT(2)/CCD   &  8x8x2     & C$_4$Si$_{12}$H$_{36}$ &&  1.985    &&    1.1     &   16.3     &    1.0     \\
\hline
UM11/CCD   &  8x8x2     & C$_4$Si$_{12}$H$_{36}$ &&  2.004    &&            &            &            \\
UB3LYP/CCD &  8x8x2     & C$_4$Si$_{12}$H$_{36}$ &&  2.077    &&            &            &            \\
PBE\footnotemark[2] & 6x6x2 &   (all atoms)      &&  2.053    &&            &            &            \\
\hline
\hline
\end{tabular}
\footnotetext[1]{Quantity computed as $\frac{|V_1-V_2|}{\frac{(V_1+V_2)}{2}}\times 100$
with $V_1$ the preceding table entry with an appropriate incrementally smaller model specification.}
\footnotetext[2]{Plane-wave calculation reported in Ref. \cite{SDEprb2016}.}
\end{table*}

When transition-metal silicon-substitutional defects are considered,
the MO structure changes significantly as compared with the V$_{\mathrm{Si}}$-type defects. 
As an example, the MOs of the chromium defect system are shown in Fig.\ \ref{fig_tmmo}.
Starting from a T$_{\mathrm{d}}$ geometry in the non-interacting limit (Fig.\ 2c),
an optimization will distort to a C$_{\mathrm{3v}}$ symmetry (Fig.\ 2b),
while also hybridizing the Cr and CH$_3$ valence orbitals (Fig.\ 2a).
It can be seen that for the initial T$_{\mathrm{d}}$ geometry (Fig.\ 2c),
when an optimization is performed on the neutral ${}^0$Cr$^{0}$ species 
all of the SOMOs can simply become doubly occupied Cr $d$ orbitals, 
and, in practice, this electronic configuration does not always 
facilitate the required orbital mixing needed for the optimization to proceed. 
For this reason we chose instead the ${}^4$Cr$^{3+}$ state,
which starts the optimization engaging the 3t$_2$ orbitals symmetrically.
Introducing the charge draws in the CH$_3$ dangling bonds and initiates orbital mixing,
while the quartet open-shell system can still be easily described 
using a single Hartree-Fock determinant.

In all QM calculations core orbitals were kept frozen,
no molecular symmetry was enforced, and a spherical harmonic basis was used.
For the SIMOMM optimizations, the MM partition was always treated using MM2 parameters \cite{mm2},
and the default maximum nuclear gradient convergence threshold was loosened
to \num{1E-3} Hartree/Bohr, since this was shown to have a relatively
small effect on the final excitation energies while reducing the
number of iterations considerably. DFT and TD-DFT calculations
were performed in GAMESS using a very tight grid (\texttt{JANS}=2).
We utilize the 6-31G, 6-31G$^*$, and 6-31+G$^*$ basis sets \cite{HPtca1973,FPHjcp1982,CCSjcc1983,KBSjcp1980,GJPcpl1992}
and the correlation-consistent basis sets of Dunning \cite{dunning1989,woon1993,dunning2001}.
Here cc-pV$X$Z and aug-cc-pV$X$Z are abbreviated as CC$X$ and ACC$X$, respectively, 
where $X$ is the cardinal number of the basis set ($X$ = D, T, Q, $\ldots$).
For vacancy defect calculations ghost functions were also included 
to improve basis set convergence. These consisted of Si functions
in the specified basis set and were placed at the vacancy-defect site.

\section{Results and Discussion}
\label{sec_randd}

The primary goal of this work is to propose and validate 
Gaussian-based approaches for generating high-accuracy
ground- and excited-state properties and energetics
of vacancy and substitutional defects in semiconductors.
While the procedures explored here are, in principle,
systematically improvable to the exact solution,
the steep computational scaling of the most accurate 
methods limits the scope of their application.
Fortunately, photoluminescence spectra are available 
for benchmarking new methods and this facilitates 
convergence tests. Much of this study is thus devoted 
to identifying for use in future studies those levels of theory 
that offer a good compromise between accuracy and computational cost.

\subsection{Defect geometry convergence using SIMOMM}
\label{sec_simomm}

As this is the first application of SIMOMM to deep-center defects, 
it is important to begin by testing whether the resulting 
geometrical parameters converge with increasing model size.
Starting from a bulk model with perfect crystal coordinates, 
introduction of a point defect followed by optimization with SIMOMM 
causes the atoms directly adjacent to the defect site to break symmetry,
as Jahn-Teller distortion elongates the primary symmetry axis \cite{PMNprb2017}.
A good single quantity to monitor for convergence is thus 
the average distance between the defect position and the four 
surrounding atoms, or $\bar{R}$. The 4H-SiC polytype was chosen 
for these tests because, unlike 3C-SiC, it has an anisotropic unit cell, 
and, unlike 6H-SiC, 4H-SiC is not too large to consider 
multiple concentric supercell dimensions (in integer-unit increments). 
While the exact geometrical structure has not been measured,
prior plane-wave DFT calculations placed $\bar{R}$ close to 2.0 \AA \cite{SDEprb2016}.
A desirable level for a convergence threshold is then a distance
$\Delta\bar{R}<0.1$ \AA, corresponding to $\Delta\bar{R}<5.0$\% in this case. 

Table \ref{tab_simomm} collects $\bar{R}$ values resulting from optimizations 
performed using various supercell sizes, QM model sizes, and levels of theory. 
When the QM model was treated at the UMBPT(2)/STO-3G level of theory
with an adequate bulk MM model supercell of 128 unit cells (8x8x2),
a rather large QM model size of C$_{4}$Si$_{12}$H$_{36}$ [Fig.\ \ref{fig_mod}(b)]
was required before reaching the desired 5\% convergence.
Since the 252-electron C$_{4}$Si$_{12}$H$_{36}$ QM model 
would be computationally intractable for many accurate QM theories,
this motivated us to investigate the effect of increasing the basis set size.
Switching from the STO-3G to the CCD basis set improved convergence with the QM model size,
and it was found that, when used with the 8x8x2 supercell,
the smallest 60-electron C$_4$H$_{12}$ QM model [Fig.\ \ref{fig_mod}(a)],
produced a $\bar{R}$ value in agreement to within 5\% 
with the best $\bar{R}$ values reported here and in Ref. \cite{SDEprb2016}.
The 8x8x2/C$_4$H$_{12}$ model treated at the UMBPT(2)/CCD level of theory
represents the best compromise of model sizes we tested for 4H-SiC.

When a larger number of atoms are required in the QM model,
DFT methods can also be used in conjunction with SIMOMM optimizations.
For 4H-SiC, when the M11 functional was used in conjunction with the CCD basis set, 
an 8x8x2 supercell, and the C$_4$Si$_{12}$H$_{36}$ QM model, 
SIMOMM optimizations produced a $\bar{R}$ value of 2.007 \AA.
This value is in agreement to within 5\% of our best UMBPT(2)/STO-3G result, 
our best UMBPT(2)/CCD result, and the literature plane-wave PBE value. 
Another popular functional choice, UB3LYP, was also tested and found to 
give a higher $\bar{R}$ value that was in good agreement with the plane-wave PBE result. 
These initial tests indicate that the comparatively inexpensive
DFT-based SIMOMM optimizations can provide accuracies 
comparable to large-basis MBPT(2) calculations, though
testing more functionals is outside the scope of this study.

Solid-state geometries used in the remainder of this work were 
optimized using SIMOMM employing the UMBPT(2)/CCD QM method
and the parameters given in Table \ref{tab_params}. 
Convergence tests were also performed on 3C-SiC, 
where improved convergence behavior was noted as compared with 4H-SiC.
For the comparatively anisotropic 6H-SiC lattice, 
we used the largest affordable roughly-cubic supercell, 
having dimensions 9x9x1.4. The 3C, 4H, and 6H polytypes 
make an interesting case study for testing our methods,
since there is varying degree of anisotropy of the unit cells 
with little other significant change in the environment of the defect.

\subsection{Charge and multiplicity of the ground-state}
\label{sec_gs}

\begin{table}
\caption{\label{tab_params} Parameters defining the computational models}
\begin{tabular}{lccc}
\hline
\hline
&\multicolumn{3}{c}{SiC polytype} \\
\cline{2-4}
                         &  3C   &  4H   &  6H     \\
\hline
space group              & F43m  & P63mc & P63mc   \\
$a$(\AA)                 & 4.368 & 3.079 & 3.079   \\
$c$(\AA)                 &  ---  & 10.07 & 15.12   \\
supercell boundaries     & 4x4x4 & 8x8x2 & 9x9x1.4 \\
MM crystal atoms         & 865   & 1561  & 1824    \\
MM hydrogen atoms        & 539   &  955  & 1007    \\
unique Si-defect sites   &  1    &   2   &   3     \\
QM crystal atoms         &  4    &   4   &   4     \\
QM hydrogen atoms        &  12   &   12  &   12    \\
\hline
\hline
\end{tabular}
\end{table}

A major challenge in the study of solid-state defects and their 
photoluminescence spectra, assuming knowledge of the material's
polytype and the defect type, is the characterization of the 
electronic ground-state of the defect site in terms of its charge and multiplicity.
One consequence of the high symmetry of point defects is orbital degeneracy, 
and, in analogy to Hund's rule for atoms, this can lead to unusual charges 
and multiplicities being the most energetically favorable.
Energy-ordering states related by incremental changes in charge and multiplicity
can be problematic using electronic structure methods such as DFT and TD-DFT
because they typically treat each case with a different SCF reference. 
Ideally, a method should instead build a series of states from the same 
correlated reference, as can be done using the EOMCC family of methods. 
When the appropriate level of correlation effects are included,
these methods will provide a highly accurate description 
of energy differences between various potential ground states.

Several possible 3C-SiC V$_{\mathrm{Si}}$ ground states are illustrated 
in Fig.\ \ref{fig_ipm}, where they are represented qualitatively using 
independent-particle-model orbital energy levels. 
By now there is consensus that the two most stable
electronic configurations are the neutral $S=1$ state [V$_{\mathrm{Si}}^{0}$(${}^{3}$A$_2$)]
and the anionic $S=\frac{3}{2}$ state [$V_{\mathrm{Si}}^{-}$(${}^{4}$A$_2$)], 
with the latter being the ground state for all three SiC polytypes.
Less is known about the relative energies of other states,
e.g., V$_{\mathrm{Si}}^{0}({}^{1}$A$_1)$, V$_{\mathrm{Si}}^{2-}$(${}^{1}$A$_1$), 
or V$_{\mathrm{Si}}^{-}$(${}^{2}$E). Since the V$_{\mathrm{Si}}^{0}$(${}^{3}$A$_2$)
species spontaneously ionizes to form the $V_{\mathrm{Si}}^{-}$(${}^{4}$A$_2$) 
species, it must be that the additional stabilizing exchange energy produced
in the anionic form is greater than the energy gained by breaking 
the symmetry of the t$_2$ orbital to form its 5e and 14a$_1$ components.

\begin{figure}
\hspace{-0.4in}\includegraphics[trim={0 0 0 3cm},clip,width=0.45\paperwidth]{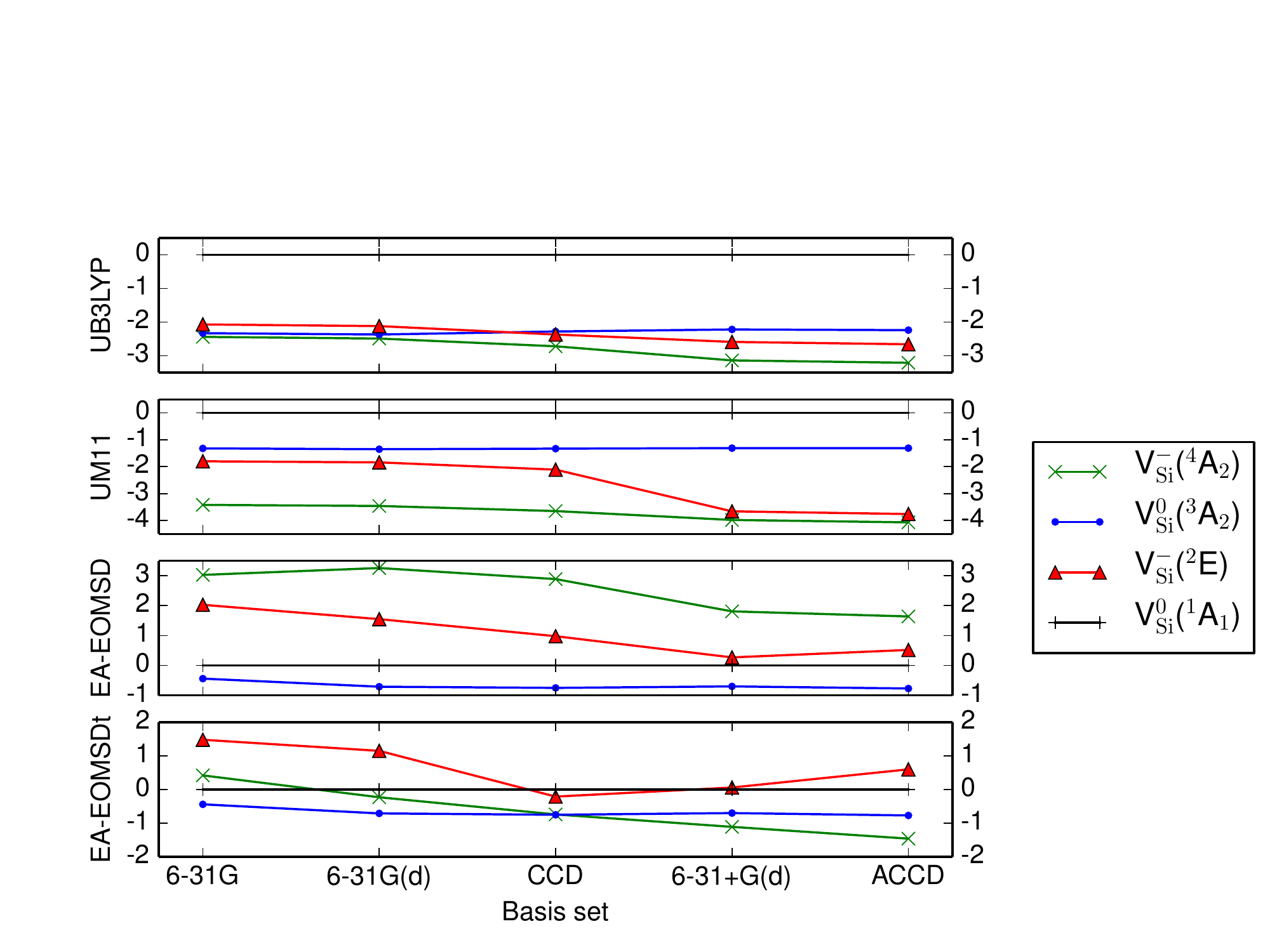}
\caption{\label{fig_states} Relative energies (in eV) 
of various electronic states of V$_{\mathrm{Si}}^-$ in 3C-SiC.
Each value is computed using the designated method and basis set
and reported with respect to the corresponding
V$_{\mathrm{Si}}^0$($^1$A$_1$) energy.}
\end{figure}

Fig.\ \ref{fig_states} plots relative energies of several low-lying
3C-SiC V$_{\mathrm{Si}}$ states as a function of basis set size 
using the UB3LYP, UM11, EA-EOMCCSD, and EA-EOMCCSDt\{3\} methods. 
Let us first consider these results in terms of what is known.
All combinations of method and basis set correctly place
the ${}^{3}$V$_{\mathrm{Si}}^0$ state below the ${}^{1}$V$_{\mathrm{Si}}^0$ state,
but there is great variation in the quantitative difference.
Beyond this, the ACCD basis set results for the UB3LYP, UM11, 
and EA-EOMCCSDt\{3\} methods also correctly place 
the ${}^{4}$V$_{\mathrm{Si}}^{-}$ state lowest.
Considering the remaining states, it is seen that the
energy-ordering provided by the DFT and EA-EOMCCSDt\{3\} methods 
differ qualitatively and further discussion is warranted.

One potentially consequential discrepency between the DFT and EA-EOMCCSDt\{3\}
state orderings is their relative placement of the anionic V$_{\mathrm{Si}}^-$(${}^2$E) state
with respect to the neutral V$_{\mathrm{Si}}^0$(${}^3$A$_2$) and V$_{\mathrm{Si}}^0$(${}^1$A$_1$) states. 
Limiting the discussion to the ACCD basis set results in Fig.\ \ref{fig_states},
the DFT methods place both anionic states lower than the neutral states,
while the EA-EOMCCSDt\{3\} method places the V$_{\mathrm{Si}}^-$(${}^2$E) state
more than 2 eV higher than the V$_{\mathrm{Si}}^-$(${}^4$A$_2$) state,
and, importantly, also above both neutral states. Experimental realization 
of a Lambda system such as the one proposed in Ref.\ \cite{SDEprb2016} based on DFT calculations,
may be compromised by the possibility of system ionization during excitation or relaxation
processes occurring between the V$_{\mathrm{Si}}^-$(${}^2$E) and V$_{\mathrm{Si}}^-$(${}^4$A$_2$) states.

Returning to comment on the basis-set dependence of the computational models,
all methods presented in Fig.\ \ref{fig_states} show a significant ($>$1 eV) 
shift in at least one of the reported states when going from the 
6-31G$^*$ to 6-31+G$^*$ basis sets. This demonstrates the importance 
of diffuse functions for the accurate energy-ordering of defect states.
Both the UB3LYP and EA-EOMCCSDt methods exhibit a basis-set dependence 
of the state ordering, with the EA-EOMCCSDt state-ordering not completely 
resolved until the ACCD basis set is employed. It is thus important 
to use good-quality basis sets with diffuse functions
when performing energy-ordering studies on minimal vacancy defect models. 

\begin{table*}
\caption{\label{tab_states} Relative energies (in eV) for 
the V$_k^{0}({}^1$A$_1)$$\rightarrow$V$_k^{-}({}^4$A$_1)$ transition, 
computed using various method and basis-set combinations.}
\begin{tabular}{ccccccccccccccccc}
\hline
\hline
 UB3LYP & UM11  &&     \multicolumn{5}{c}{EA-EOMCCSD} 
                &&  \multicolumn{6}{c}{EA-EOMCCSDt\{3\}} && GW approx.\footnotemark[1] \\
 ACCT   & ACCT  && 6-31G&6-31G$^*$& CCD &6-31+G$^*$&ACCD  
                && 6-31G&6-31G$^*$& CCD &6-31+G$^*$&ACCD  &ACCT  && plane-wave\\
\hline 
 -3.28  & -4.07 && 3.03 &   3.26  & 2.89&    1.81  & 1.64    
                && 0.42 &  -0.24  &-0.75&   -1.11  &-1.46 &-1.67 &&-1.58     \\
\hline
\hline
\end{tabular}
\footnotetext[1]{Ref.\ \cite{BRprb2011}; Literature computational values were obtained using the $GW$ approximation.}
\end{table*}

Without benchmark values for comparison, it is difficult to draw definitive conclusions
from the data in Fig.\ \ref{fig_states} about the relative accuracy of these methods.
Table \ref{tab_states} provides a quantitative comparison of computed energy differences 
for the V$_k^{0}$(${}^1$A$_1$)$\rightarrow$V$_k^{-}$(${}^4$A$_1$) transition, with a literature 
plane-wave-based GW-approximation value also included for comparison \cite{BRprb2011}.
The UB3LYP and UM11 DFT approaches produce relative energies over twice as large as the GW approximation,
while the EA-EOMCCSD method consistently produces the wrong sign for the energy difference.
The EA-EOMCCSDt\{3\} method fares much better. When the ACCD and ACCT basis sets are employed,
EA-EOMCCSDt\{3\} produces values differing by only $\sim$~0.1 eV from the $GW$ approximation.
This provides supportive evidence that the EA-EOMCCSDt\{3\} produces the most accurate 
relative energetics of the four methods used here, and thus it likely also provides
the most reliable state-ordering in Fig.\ \ref{fig_states}.

\subsection{Basis set convergence of excitation energies}
\label{sec_es}

In this section we investigate the accuracy and basis-set convergence 
of excitation energies produced out of the V$_k^{-}$(${}^4$A$_1$) state
using EOMCC and TD-DFT methods.
In Ref.\ \cite{SDEprb2016} plane-wave DFT calculations were used to 
qualitatively order a series of doublet and quartet excited states, 
with symmetries predicted using a purely group-theoretic approach. 
It is thus an interesting question whether our Gaussian-based procedure
will produce energy-ordering of excited states similar to the plane-wave DFT calculations.
Before making such comparisons, in this section we establish an appropriate 
method and basis set for our approach through convergence tests.
 
Table \ref{tab_es} collects excitation energies generated using various methods and basis sets,  
with only the two lowest-lying quartet states, 1${}^4$A$_1$ and 1${}^4$E, reported.
In terms of the basis set convergence, it is clear from Table \ref{tab_es} that,
regardless of the method, diffuse functions are essential to the accuracy of the model.
When the 6-31+G$^*$ and ACCD basis sets including diffuse functions are employed, 
the resulting excitation energies are within 0.25 eV of the corresponding ACCT results,
providing a practical alternative to ACCT in defect calculations where expense is a limiting factor.
Full IP- and EA-EOMCCSDT results are also included for the 6-31G basis set;
the strong similarity of the values produced by the active-space methods and
their parent methods (within 0.01 eV) indicates that the active-space orbitals 
are an appropriate set for capturing the most important triples effects.

\begin{table}
\caption{\label{tab_es} Convergence of 3C-SiC excitation energies (in eV) 
for transitions from the V$_{\mathrm{Si}}^{-1}$(X${}^4$A$_1$) ground state 
to the excited ${}^4$A$_1$ state (above) and ${}^4$E state (below), corresponding
to orbital transitions dominated by 14a$_1$$\rightarrow$15a$_1$ and 14a$_1$$\rightarrow$6e
character, respectively [Fig.\ \ref{fig_ipm}(a)].}
\begin{tabular}{llcccccc}
\hline
\hline
              &     \multicolumn{6}{c}{Basis set}            \\
Method        & 6-31G&6-31G$^*$& CCD &6-31+G$^*$&ACCD & ACCT \\
\hline
UB3LYP        &  2.50 &  2.48  & 2.35&  1.89   & 1.81 & 1.76 \\
\hline
IP-EOMSD      &  7.56 &  7.33  & 6.44&  3.85   & 2.49 & 2.62 \\
IP-EOMSDt     &  7.61 &  6.10  & 5.07&  2.97   & 2.88 & 2.91 \\
IP-EOMSDT     &  5.78 &        &     &         &      &      \\
\hline
EA-EOMSD      &  2.66 &  2.60  & 2.22&  0.82   & 0.48 & 0.31 \\
EA-EOMSDt     &  2.60 &  2.56  & 2.43&  3.00   & 2.94 & 2.75 \\
EA-EOMSDT     &  2.60 &        &     &         &      &      \\
\hline
\hline
UB3LYP        &  2.51 &  2.47  & 2.34&  1.88   & 1.80 & 1.76 \\
\hline
IP-EOMSD      &  2.44 &  2.37  & 2.29&  1.96   & 3.27 & 3.27 \\
IP-EOMSDt     &  2.54 &  2.47  & 2.34&  1.79   & 3.47 & 3.36 \\
IP-EOMSDT     &  2.54 &        &     &         &      &      \\
\hline
EA-EOMSD      &  3.04 &  3.09  & 2.23&  0.22   & 0.06 & 0.14 \\
EA-EOMSDt     &  3.74 &  4.09  & 3.34&  1.38   & 1.32 & 1.35 \\
EA-EOMSDT     &  3.73 &        &     &         &      &      \\
\hline
\hline
\end{tabular}
\end{table}

Considering more closely the X${}^4$A$_1\rightarrow 1{}^4$A$_1$ transition,
in Table \ref{tab_es} a significant discrepancy is found between 
the excitation energies produced by the UB3LYP, EA-EOMCCSD, 
and EA-EOMCCSDt\{3\} methods. Differences between EA-EOMCCSD 
and EA-EOMCCSDt are attributable to the significant contributions 
from $r_{\mu,3p-2h}$ amplitudes (see Eqs.\ \ref{eq_rea} and \ref{eq_as}) 
found for the X${}^4$A$_1$ state. The EA-EOMCCSDt\{3\} and UB3LYP methods
are also in disagreememnt for the same transition by nearly 1.0 eV. 
The EA-EOMCCSDt\{3\} method places the $1{}^4$A$_1$ state 1.4 eV higher 
in energy than the $1{}^4$E state, while UB3LYP predicts the two excited states 
to be quasi-degenerate. Since a well-known deficiency of TD-DFT is that
it does not incorporate two-electron transitions, this can again be 
attributed to the significant $r_{\mu,3p-2h}$ amplitudes appearing
in the EA-EOMCCSDt\{3\} calculations, which indicate that the
excitation is not a pure one-electron transition. Indeed, 
the UB3LYP 1${}^4$A$_1$ configuration state function is dominated by
one large ($>0.98$) amplitude out of the X${}^4$A$_1$ state
with all other amplitudes being small ($<0.1$), indicating 
that there are virtually no accompanying orbital rotations.

Table \ref{tab_es} also includes IP-EOMCC results, as these 
are often more accurate than the EA-EOMCC methods if the target
radical anionic ($N+1$)-electron wave function more closely resembles 
a doubly anionic ($N$+2)-electron species rather than the $N$-electron one.
The EA-EOMCCSDt\{3\} and IP-EOMCCSDt\{6\} results converge toward a similar value
for the 1${}^4$A$_1$ state, but the IP-EOMCCSD and IP-EOMCCSDt\{6\} results
do not converge systematically for the X${}^4$A$_1\rightarrow 1{}^4$E transition.
In other situations the IP-EOMCC methods may be a better choice,
but since the EA-EOMCC methods are a more convenient and accurate 
choice for these systems we focus on them here for the remainder of this study.

\subsection{Benchmarking excitation energies of silicon-vacancy defects in 4H- and 6H-SiC}
\label{sec_poly}

Photoluminescence spectra have previously been obtained for 4H- and 6H-SiC
and these can be used to benchmark the accuracy of our approach, which so
far has been tested only on 3C-SiC. In Table \ref{tab_poly} excitation energies 
computed with the UB3LYP/ACCT and EA-EOMCCSDt\{3\}/ACCD methods are compared 
with related photoluminescence measurements for all V$_{\mathrm{Si}}^-$ defect types 
in 4H- and 6H-SiC. The EA-EOMCCSDt\{3\} computational values for the
X${}^4$A$_1 \rightarrow 1^2$E transition are all within 0.1 eV to the measured values.
The EA-EOMCCSDt\{3\} energy-ordering of different defect types within a given polytype
also qualitatively matches with measurements, indicating this method may be helpful 
in future studies for distinguishing defect types differing subtly in energy. 
For both the 4H and 6H polytypes the X${}^4$A$_1 \rightarrow 1{}^4$A$_1$ transition
is nearly 3 eV, which supports the similar assignment made for 3C-SiC in Table \ref{tab_es}.
We note that the magnitude of the error increases with increasing unit-cell anisotropy,
and thus the larger errors f 6H-SiC would likely be reduced by utilizing a 
more complete supercell during the SIMOMM optimization.  

Comparing instead the TD-DFT calculations with the measured values, 
somewhat erratic UB3LYP results were found for the same set of geometries.
In more than one case the energies are too large by over 0.5 eV when compared 
to the corresponding benchmark values, and in almost all cases the $1{}^4$A$_1$ and $1{}^4$E
states lie very close in energy, similar to what was found for 3C-SiC in Sect.\ \ref{sec_es}. 
For the k2-type 6H-SiC defect, where the X${}^4$A$_1 \rightarrow 1{}^4$A$_1$
excitation energy is too small by more than 1 eV, the underlying DFT calculation
has presumably converged to the 1${}^4$A$_1$ state, as evidenced by it being
nearly degenerate with the $1{}^4$E state.
Since our goal was simply to identify the most accurate methods for our procedure,
we did not attempt to rotate the KS orbitals in pursuit of a lower-energy state.

\begin{table}
\caption{\label{tab_poly} Comparison of computed and measured 
excitation energies (in eV) for transitions out of the
$V_{\mathrm{Si}}^{-}$(${}^4$A$_1$) state in 4H- and 6H-SiC.
The upper and lower tables differ only in the computational 
method used to generate vertical excitation energies, as indicated,
while the final line provides measured values for reference.}
\begin{tabular}{lcccccc}
\hline
\hline
                                &  \multicolumn{6}{c}{UB3LYP/ACCT}                       \\
\cline{2-7}
                                &  \multicolumn{2}{c}{4H-SiC}&&\multicolumn{3}{c}{6H-SiC}\\
\cline{2-3}\cline{5-7}
State                           & k(V1)             & h(V2) &&  k1(V1) & h(V2)  & k2(V3) \\
\hline
V$_{\mathrm{Si}}^{-}$($1{}^4$A$_1$)&1.990            &  1.513 &&  2.051  & 1.754  & 1.218   \\
V$_{\mathrm{Si}}^{-}$($1{}^4$E)  &  1.982            &  1.497 &&  2.051  & 1.754  & 0.061   \\
\hline
\hline
                               &  \multicolumn{6}{c}{EA-EOMCCSDt/ACCD}                   \\
\cline{2-7}
                               &  \multicolumn{2}{c}{4H-SiC}&&\multicolumn{3}{c}{6H-SiC} \\
\cline{2-3}\cline{5-7}
State                           & k(V1)             & h(V2) &&  k1(V1) & h(V2)  & k2(V3) \\
\hline
V$_{\mathrm{Si}}^{-}$($1{}^4$A$_1$) & 2.968             & 3.084   &&  2.966  & 2.967  & 2.961   \\
V$_{\mathrm{Si}}^{-}$($1{}^4$E)   & 1.424             & 1.321 &&  1.334  & 1.331  & 1.329   \\
\hline
\hline
Experiment\footnotemark[1]      & 1.438            & 1.352 &&  1.433  & 1.398  & 1.368   \\
\hline
\hline
\end{tabular}
\footnotetext[1]{Photoluminescence measurements of the $X{}^4A_1 \rightarrow 1{}^4E$ transition taken from Refs. \cite{SSCprb2000} and \cite{WMCprb2000}}
\end{table}

\subsection{Chromium silicon-substitutional defects in SiC}
\label{sec_tm}

Photoluminescence frequencies of the V$_{\mathrm{Si}}^-$ defect 
are unsuitable for leveraging existing telecommunication technology,
and there is consequently ramping interest in screening transition-metal defects 
for a color center with an emission frequency compatible with fiber-optic technology.
While many methods struggle to accurately describe transition-metal excitation energies, 
the active-space EA- and IP-EOMCC methods have recently proven to be 
very successful for transition metals when used appropriately \cite{PPsilver,PPgold}.
As a more challenging test of our approach, here we make a first attempt 
at reproducing the excitation energy for a transition-metal defect in SiC.
The photoluminescence spectra for a single chromium defect in SiC has been recently measured,
and the authors of Ref.\ \cite{KDWprb2017} have reported peaks at 1.1587 and 1.1898 eV 
${}^3$Cr$^{4+}$ defect corresponding to the $h$- and $k$-type silicon sites of 4H-SiC, respectively. 

After obtaining a converged quartet Cr$^{3+}$(CH$_{3}$)$_4$ geometry for 3C-SiC, 
as described in Sect.\ \ref{sec_comp}, the preferred charge and multiplicitly 
of the ground state was investigated using EE-EOMCCSD and EA-EOMCCSDt calculations. 
Our initial exploratory calculations were performed using the 3C polytype of SiC
because we encountered convergence problems for Cr-embedded 4H-SiC.
From the results presented in Table \ref{tab_cr} it can be seen that calculations 
performed at all reported basis set levels place the Cr$_{\mathrm{Si}}^{0 }({}^3$A$_2)$ 
species lowest in energy. In this case there is no change 
in the energy-ordering of states with increasing basis set, and, as in Sect.\ \ref{sec_gs}, 
ground-state energy differences computed using the ACCD basis set appear adequately converged.
This agrees with the ground-state multiplicity predicted in Ref.\ \cite{KDWprb2017}, 
but the oxidation state differs from the 4+ oxidation state reported there
(presumably their value corresponds to the oxidation number of the source material).
A Mulliken population analysis confirms the predicted oxidation state is close to zero, 
producing a value of 0.15 a.u.\ on the Cr atom when the ACCD basis set is used.

In Ref.\ \cite{KDWprb2017} the authors posited that the observed 4H-SiC Cr$_{\mathrm{Si}}^{0 }$
transition is due to a X${}^3$A$_2 \rightarrow 1{}^1$A$_1$ transition. Our 3C-SiC result
for that transition is 1.15 eV, in good agreement with the measured 4H-SiC values.
Of further interest are the result of our calculation for the 3C-SiC Cr$_{\mathrm{Si}}^{0 }$
X${}^3$A$_2 \rightarrow 1{}^3$A$_2$ transition, which yielded a value of 1.44 eV.
This transition is close enough to the fiberoptic C-band that it may be worth 
further consideration, especially since these defects can already be reliably 
created and measured.

\begin{table}
\caption{\label{tab_cr} Relative energies of chromium-defect states of incremental charge and multiplicity.
Energies were computed using the EE-EOMCCSD/ACCD and EA-EOMCCSDt/ACCD method for systems with an even and odd 
numbers of electrons, respectively. 
All values are reported relative to the neutral singlet Cr$_{\mathrm{Si}}^{0 }({}^1A_1)$ state, in eV.}
\begin{tabular}{lccc}
\hline
\hline
Species(state)                    &  6-31G &6-31+G$^*$&  ACCD  \\
\hline
Cr$_{\mathrm{Si}}^{2-}({}^1$A$_1)$ &  6.92  &   4.04   &  N/C   \\
Cr$_{\mathrm{Si}}^{ -}({}^2$E$  )$ &  0.11  &  -0.49   &  0.27  \\
Cr$_{\mathrm{Si}}^{0 }({}^3$A$_2)$ & -1.37  &  -1.24   & -1.15  \\
Cr$_{\mathrm{Si}}^{ +}({}^2$E$  )$ &  4.86  &   5.24   &  5.49  \\
Cr$_{\mathrm{Si}}^{ +}({}^4$E$  )$ &  6.97  &   7.47   &  7.81  \\
Cr$_{\mathrm{Si}}^{2+}({}^1$A$_1)$ & 19.01  &  19.39   & 19.63  \\
Cr$_{\mathrm{Si}}^{2+}({}^3$A$_2)$ & 19.27  &  19.77   & 20.03  \\
Cr$_{\mathrm{Si}}^{3+}({}^2$E$  )$ & 38.55  &   N/C\footnotemark[1]    &  N/C\footnotemark[1]   \\
Cr$_{\mathrm{Si}}^{4+}({}^1$A$_1)$ & 65.07  &   N/C\footnotemark[1]    &  N/C\footnotemark[1]   \\
Cr$_{\mathrm{Si}}^{4+}({}^3$A$_2)$ & 64.40  &   N/C\footnotemark[1]    &  N/C\footnotemark[1]   \\
\hline
\hline
\end{tabular}
\footnotetext[1]{The calculation did not converge.}
\end{table}

\section{Conclusions}
\label{sec_conc}

In this study we proposed and validated an {\it ab initio} Gaussian-based method for 
predicting the structure and emission frequencies of deep-center defects in semiconductors. 
The procedure is as follows: starting from perfect crystalline lattice coordinates,
the defect is introduced and the positions of the surrounding atoms are optimized 
using the QM/MM method SIMOMM.
Excitation energies are then computed by applying highly-accurate EOMCC-based methods
to a model structure consisting of several atoms immediately adjacent to the defect,
in their SIMOMM-optimized positions. While these minimal model geometries were 
sufficient to produce excitation energies comparable to the corresponding 
photoluminescence measurements, it should also be emphasized that the 
steep expense of EOMCC methods are being overcome, both through 
massively-parallel computing algorithms \cite{LFPjcp2008,BKHcpl2011}
and orbital localization schemes \cite{FBjcp2004,ANjcp2006,LPGjcp2009,EBEjctc2015,LNjctc2015}.
After breaking free of the associated intractable computational scalings,
the systematically improvable nature inherent to our SIMOMM-based method
will be a critical advantage over plane-wave methods.

It was demonstrated through convergence tests that the Gaussian-based QM/MM method SIMOMM
could achieve a similar level of accuracy to plane-wave based PBE calculations 
using around 1000 atoms in the bulk MM model. With the QM portion sufficiently constrained,
and assuming that an adequately large basis set was employed, 
both MBPT(2) and DFT with the M11 functional were shown to provide accurate geometries
with a QM treatment of only the four carbon atoms immediately adjacent to the defect center.
Given as a starting point these accurate optimized geometries, EOMCC-based methods
were shown to be powerful tools for the prediction of the electronic structure
of defect centers. Using a sufficiently large basis set,
the EA-EOMCCSDt method reliably predicted the ground state for silicon-vacancy 
defects among several states varying in charge and multiplicity,
and it produced quantitative excitation energies, always in agreement
with photoluminescence measurements to within 0.1 eV.

After establishing the accuracy of this procedure on silicon-vacancies in SiC,
a first attempt was made to apply it to a chromium silicon-substitutional defect
and EOMCC-based methods were successful there, too. For 3C-SiC, EE-EOM-CCSD was able to 
correctly predict a triplet ground state and a related excitation energy 
closely comparable to the recently measured 4H-SiC photoluminescence spectrum. 
Our calculations predicted the chromium ground-state to have a zero oxidation number however, 
in disagreement with Ref.\ \cite{KDWprb2017} which assumed a +4 Cr oxidation state.

The computational procedure developed here will
facilitate efficient screening of defect emission frequencies 
that would otherwise take years to create and measure in the laboratory. 
This method is broadly applicable to various defects in SiC
and other semiconductors, and we will use it in a subsequent
study to screen many candidate defects, including transition-metal 
substitutional defects other than Cr, in pursuit of one that emits 
in a region compatible with the exisiting fiber-optic infrastructure.
Fabrication of such a device would go a long way toward establishing
the silicon-photonic route as the leading candidate platform 
for the realization of quantum information networks.

\section{Acknowledgements}

The views expressed in this work are those of the authors and 
do not reflect the official policy or position of the United
States Air Force, Department of Defense, or the United States
Government. The DoD High Performance Computing Modernization
(HPCMO) Program and the AFRL Supercomputing Resource Center (DSRC)
are gratefully acknowledged for financial resources and computer
time and helpful support. This work was sponsored by a HASI grant
from HPCMO and a joint AFRL/AFIT grant 
awarded at the AFRL Director's authority. 
This project was enabled in part by an appointment
to the Internship/Research Participation Program at the Air Force
Institute of Technology, administered by the Oak Ridge 
Institute for Science and Education through an interagency 
agreement between the U.S. Department of Energy and EPA. 

\bibliography{library}

\end{document}